\documentclass[final,5p,twocolumn,times]{elsarticle}
\makeatletter
\def\ps@pprintTitle{%
	\let\@oddhead\@empty
	\let\@evenhead\@empty
	\def\@oddfoot{}%
	\let\@evenfoot\@oddfoot}
\makeatother
\usepackage{fancyhdr}
\usepackage{pgfbaseimage}
\pgfdeclareimage[height=0.75cm]{cc-license}{by-nc-nd}
\fancypagestyle{pprintTitle}{%
	\fancyhf{}
	\lfoot{\href{https://doi.org/10.1016/j.sna.2022.113585}{\texttt{doi:10.1016/j.sna.2022.113585}} \\ © 2022. This manuscript version is made available under a \href{https://creativecommons.org/licenses/by-nc-nd/4.0/}{CC-BY-NC-ND 4.0 license}.}%
	\rfoot{\vspace{3mm}\href{https://creativecommons.org/licenses/by-nc-nd/4.0/}{\pgfuseimage{cc-license}}}

}

\usepackage{cmap} 
\usepackage[T1]{fontenc}
\usepackage{textcomp}
\usepackage{url}
\usepackage[colorlinks,bookmarksopen,bookmarksnumbered]{hyperref}
\usepackage{siunitx}
\usepackage{subcaption} 
\usepackage{booktabs} 
\usepackage{mathtools}
\usepackage{csquotes}
\usepackage{physics}
\usepackage{array}
\usepackage{multirow}
\usepackage{makecell}
\usepackage{cancel}
\usepackage{xcolor}
\colorlet{highlight}{cyan!50}
\usepackage{soul}
\sethlcolor{highlight}
\DeclareMathOperator{\atantwo}{atan2}

\graphicspath{{./graphics/}}

\newcommand\tab[1][1cm]{\hspace*{#1}}
\newcolumntype{L}[1]{>{\raggedright\let\newline\\\arraybackslash\hspace{0pt}}m{#1}}
\newcolumntype{C}[1]{>{\centering\let\newline\\\arraybackslash\hspace{0pt}}m{#1}}

\begin{document}
	\begin{frontmatter}
		\title{Harmonic analysis of the arctangent function regarding the angular error introduced by superimposed Fourier series for application in sine/cosine angle encoders}
		
		\author[boschaddress,hhnaddress]{Stefan Kuntz\corref{cor1}}
		\cortext[cor1]{Corresponding author}
		\ead{stefan.kuntz@de.bosch.com}
		\author[boschaddress,tudresden]{Robert Dauth}
		\ead{robertalexander.dauth@bosch.com}
		\author[tudresden]{Gerald Gerlach}
		\author[hhnaddress]{Peter Ott}
		\author[boschaddress]{Sina Fella}
		\address[boschaddress]{Robert Bosch GmbH, Robert-Bosch-Allee 1, 74232 Abstatt, Germany}
		\address[tudresden]{Technische Universität Dresden, Helmholtzstraße 10, 01069 Dresden, Germany}
		\address[hhnaddress]{Heilbronn University, Max-Planck-Str. 39, 74081 Heilbronn, Germany}
		
		\begin{abstract}
			We present a rigorous analytical method for harmonic analysis of the angular error of rotary and linear encoders with sine/cosine output signals in quadrature that are distorted by superimposed Fourier series. To calculate the angle from measured sine and cosine encoder channels in quadrature, the arctangent function is commonly used.
			The hence non-linear relation between raw signals and calculated angle---often thought of as a black box---complicates the estimation of the angular error and its harmonic decomposition.
			
			By means of a Taylor series expansion of the harmonic amplitudes, our method allows for quantification of the impact of harmonic signal distortions on the angular error in terms of harmonic order, magnitude and phase, including an upper bound on the remaining error term---without numerical evaluation of the arctangent function. The same approximation is achieved with an intuitive geometric approximation in the complex plane, validating the results. Additionally, interaction effects between harmonics in the signals are considered by higher-order Taylor expansion. The approximations show an excellent agreement with the exact calculation in numerical examples even in case of large distortion amplitudes, leading to practicable estimates for the angular error decomposition.
			
		\end{abstract}
		\begin{keyword}
			Angle encoders, Rotary encoders, Spatial harmonics, Angular error harmonics, Arctangent Fourier series argument, Arctangent Taylor series
		\end{keyword}
		
	\end{frontmatter}
    \section{Introduction}
	    Angle encoders are ubiquitous and indispensable in scientific or industrial apparatuses. Common applications include servo motor controls, robotics or virtually any kind of positioning application. Many different measurement methods exist and new principles are still contrived, though most of the sensors share one common denominator: The angle is not measured directly, but rather encoded in two orthogonal signals---commonly known as \emph{sine} and \emph{cosine} channels in quadrature---that form a vector in the complex plane. The actual angle is then calculated with the arctangent function, essentially converting Cartesian coordinates to a polar angle.
	    
	    Since the angle is calculated and not directly measured, a distinction must be made between the signal and the angle domain. It has been observed that harmonic disturbances in the signal-domain---with a specific order---also lead to harmonics in the angle domain, but of different orders. This raises the question of how it is possible to estimate the angular error based on known disturbances in the signal domain. This knowledge is important to predict and potentially compensate the impact of systematic periodic errors in an encoder design. The cause of these kinds of errors depend on the measurement principle and are most commonly caused by manufacturing tolerances and imperfections in the sensor design, e.g. non-ideal magnetization and flux distribution in case of magnetic sensors.
	    The following methods are equally applicable to linear position encoder with sine/cosine output signals which are then converted to an equivalent linear distance.
	    
	  	In the existing literature, \citeauthor{Hanselman1990} gives an overview of non-ideal encoder signals and their effect on the angular error for inductive resolvers \cite{Hanselman1990,Hanselman1991,Hanselman1990a}. The effect of amplitude imbalance, phase mismatch and various cases with different harmonic distortions are shown. However, the observations contain only first-order approximations of the angular error and focus mainly on special cases encountered in inductive resolvers, which may not be applicable for other types of encoder.
	  	
	  	\citeauthor{Hou2019} conduct an empirical analysis of angular errors in capacitive encoders \cite{Hou2019}. Primarily, different error sources of the measurement principle and their impact on the harmonic composition of the angular error are discussed.  
	  	
	  	\citeauthor{Secrest2015} describe an online correction method of the angular error based on model reference-adaptive systems techniques \cite{Secrest2015}.
	  	The discussion is focused on magneto-resistive sensors and it is described that disturbance harmonics with the electrical order $h$ are introducing mechanical harmonics in the angular domain with the order of $2(h-1)$, for a sensor with periodicity $p=2$. As our results will show, this is not generally true---it is only applicable for orthogonal disturbance harmonics with equal amplitudes $A_n=B_n$ (see Eq.\,\eqref{eq:signals-fourier-definition}).
	  	
\begin{figure}
	    	\centering
	    	\includegraphics[width=\columnwidth]{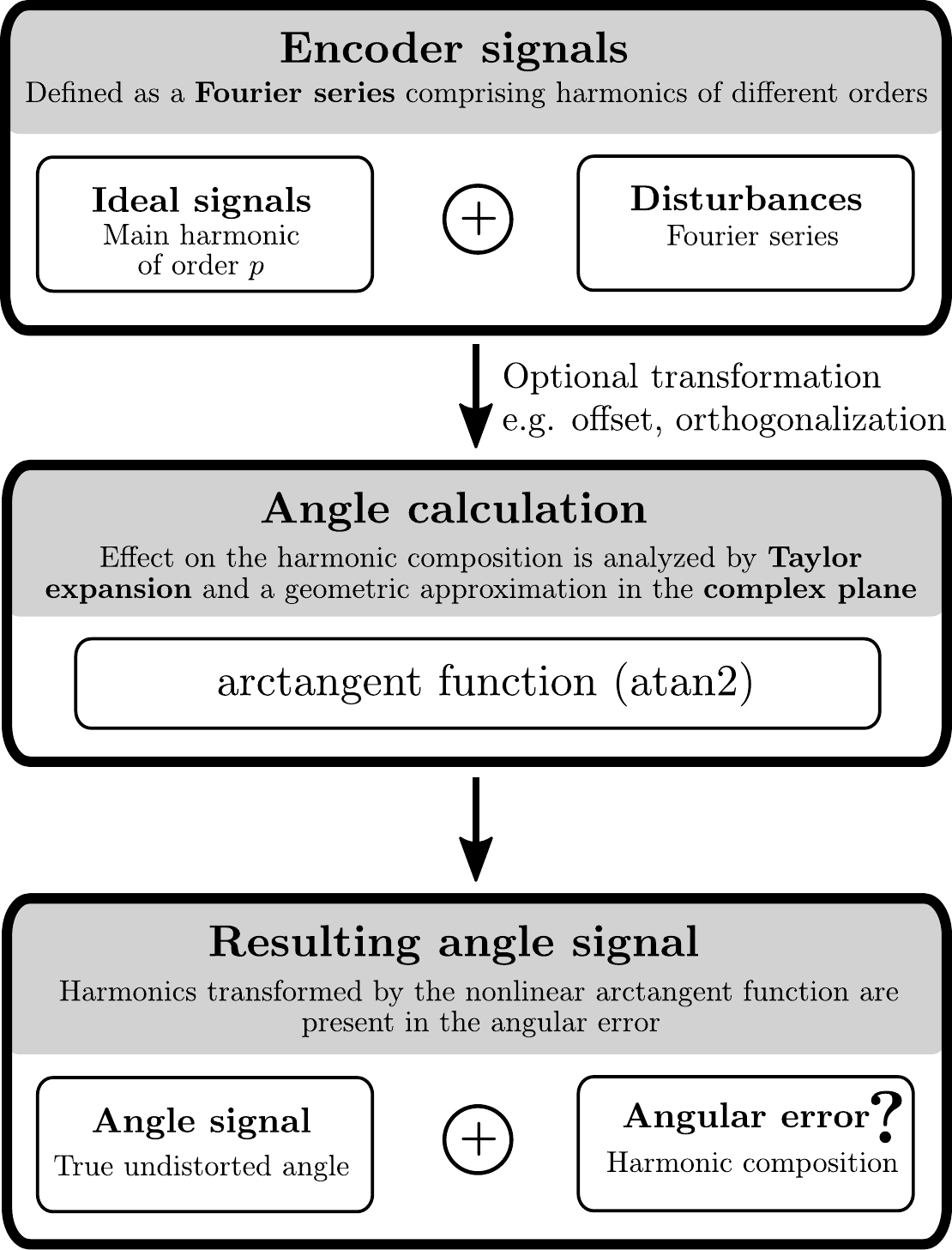}
	    	\caption{Overview of the signal flow from encoder signals to calculated angle. In practice, signals contain imperfections in the form of harmonic disturbances. The non-linear arctangent function complicates the prediction and assessment of the harmonic composition of the angular error. The methods presented in this paper enable the harmonic analysis of the arctangent function and approximation of the resulting harmonic angle error spectrum in dependence of the input disturbances in the signal domain.}
	    	\label{fig:introduction}
	    \end{figure}

	    However, a generally applicable, in-depth analysis of angular error harmonics introduced by periodic disturbances in the signal domain of angle encoders {(\autoref{fig:introduction})} is apparently missing from existing literature. In the following we will present two methods to approximate the angular error without numerically evaluating the arctangent function, which allows for symbolic calculations without resorting to numerical methods.
	    
	    The motivation of this paper is to establish a deeper understanding of
	    the nonlinearity of angle encoders and to enable a thorough analysis
	    of the cause of observed harmonics in various encoder designs. The proposed method should be a valuable tool in the future development of advanced error compensation algorithms and encoder models.
	   
	    \begin{figure}
	    	\centering
	    	\includegraphics[width=\columnwidth]{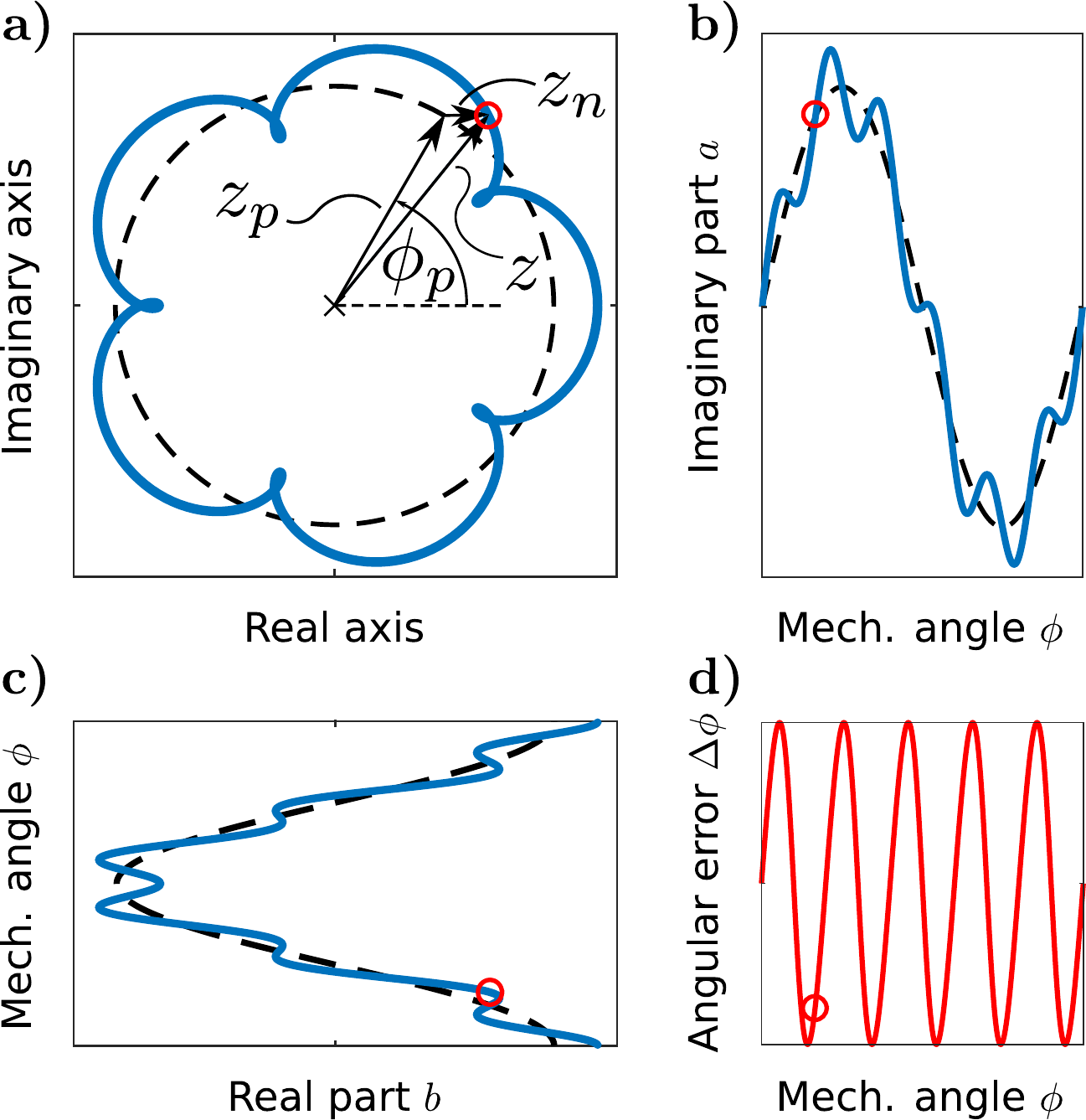}
	    	\caption{Example scenario for a sensor with $p=1$ and one imposed disturbing harmonic $z_n$ with order $n=6$. Due to the periodicity follows for the \emph{mechanical} angle $\phi=\phi_p$. The vector of the main harmonic $z_p$ lies on the unit circle (dashed line).
    		a) Lissajous figure with corresponding vectors for $\phi=60^\circ$, b) imaginary part of the composed signal, c) real part of the composed signal, d) angular error which is introduced by $z_n$.}
	    	\label{fig:vectors-complex-plane}
	    \end{figure} 
	    
   		\subsection{Definition of encoder signals}
		Let $a$ and $b$ be the output signals from an angle encoder with periodicity $p$. By considering only periodic disturbances, the encoder signals can be written as a Fourier series. In the general case the signals are then defined as
			\begin{align} \label{eq:signals-fourier-definition}
				\begin{split}
				a(\phi) &= A_0 + 	\underbrace{A_p\sin(p\phi+\theta_p)}_\text{main harmonic of $a$} + \sum_{\substack{n=1 \\ n\neq p}}^{\infty} A_n\sin(n\phi+\theta_n), \\
				b(\phi) &= B_0 + 		\underbrace{B_p\cos(p\phi+\psi_p)}_\text{main harmonic of $b$} + \sum_{\substack{n=1 \\ n\neq p}}^{\infty} B_n\cos(n\phi+\psi_n)
				\end{split}
			\end{align}
		   	where the main harmonic\footnote{We avoid the term \emph{fundamental} harmonic because it could be ambiguous when $p>1$.} of order $p$ is the usable sensor signal which is utilized to encode and later calculate the angle $\phi$. All amplitudes are defined to be positive or zero to avoid an ambiguity with the corresponding phase shift.
		   	
		   	The remaining terms are the DC offsets ($A_0$, $B_0$) and a sum of $n$\textsuperscript{th} order harmonics. They represent undesired disturbances that will cause an angular error. In addition, the main harmonic may also contain errors in form of an amplitude mismatch of $A_p$ and $B_p$ as well as an orthogonality error $\theta_p$ and $\psi_p$ which can be understood as the deviation from ideal quadrature.
		   	For simplicity we initially assume
		   	\begin{align}
		   		\label{eq:signal-assumptions}
			   	\begin{split}
			   	A_0 &= B_0 = 0,\\
			   	A_p &= B_p = 1,\\
			   	\theta_p &= \psi_p = 0
			   	\end{split}
		   	\end{align}
		   	for the following analysis, so that only the ideal terms
		   	\begin{align}
			   	\cos(p\phi)
			   	\shortintertext{and}
			   	\sin(p\phi)
		   	\end{align} 
		   	remain as the main harmonic together with higher-order harmonics and possibly lower-order harmonics if $p>1$. 
		   	While this is initially restrictive, these conditions can be relaxed later on by transformation to an equivalent harmonic as discussed in \autoref{sec:special-cases}.
		   	
		   	Instead of treating the encoder signals $a(\phi)$ and $b(\phi)$ separately, they can also be considered together as an angle-dependent \emph{vector} in the complex plane (\autoref{fig:vectors-complex-plane}) as
		   	\begin{equation}
		   		\label{eq:signal-definition}
			   	z = b + ia = z_0 + z_p + \sum_{\substack{n=1 \\ n\neq p}}^{\infty} z_n, 
		   	\end{equation}
			where $i$ is the imaginary unit.\footnote{Although mathematically imprecise, we make no distinction between $\mathbb{C}$ and the set of two-dimensional vectors $\mathbb{R}^2$.} Note that the definition of encoder signals that is used here is also applicable to multi-phase systems with more than two signals, after appropriate transform (e.g. Clarke) to the complex plane.
			
			Note that for a periodicity of $p>1$, a distinction must be made between the \emph{mechanical} angle $\phi$ and the \emph{electrical} angle $\phi_p$ which differs from the mechanical angle by a factor of $p$  	
		   	\begin{equation}
			   	\label{eq:mech-electr-angle}
			   	\phi_p = p\phi\,.
		   	\end{equation}
		   	If $p>1$, the mechanical angle $\phi$ can generally not be recovered unambiguously based on only the electrical angle. Additional techniques, such as e.g. counting turns or utilizing the Vernier principle together with another encoder may be used if the absolute mechanical angle is required.
		   	
\begin{figure}
		   		\centering
		   		\includegraphics[width=\columnwidth]{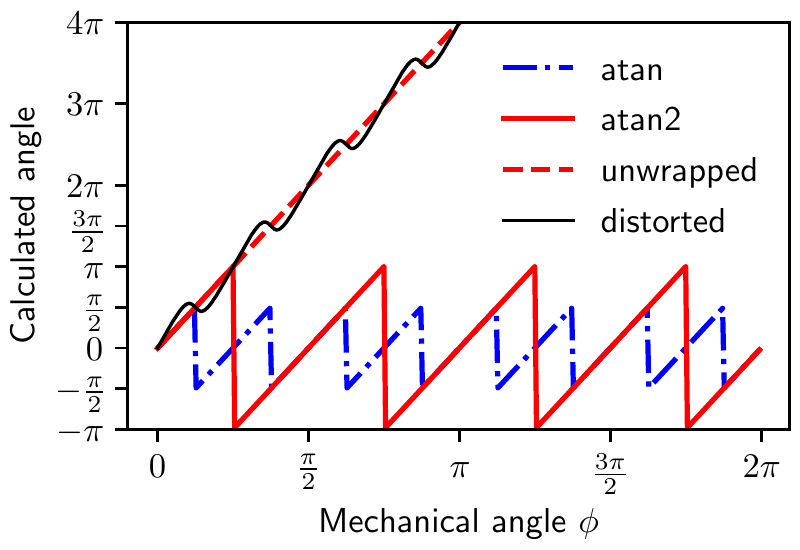}
		   		\caption{Calculated encoder angle in dependence of the mechanical reference angle $\phi$ for a sensor with periodicity $p=4$. For comparison, the angle is calculated by $\atan$ (Eq. (\ref{eq:atan})), $\atantwo$, and unwrapped $\atantwo$ functions. An exaggerated angular error deviating from the ideal output is shown as an example.}
		   		\label{fig:atan2-vs-atan}
		   	\end{figure}

		   	To calculate the encoded angle from $a$ and $b$, the arctangent function is commonly used. In case of ideal, undistorted signals
		   	\begin{equation}
		   		\label{eq:atan}
		   		\atan\left(\frac{a(\phi)}{b(\phi)}\right) = \atan\left(\frac{\sin(p\phi)}{\cos(p\phi)}\right) = \atan\left(\tan(p\phi)\right) = \phi_p \,.
		   	\end{equation}
	   		Eq. (\ref{eq:atan}) is only valid for a restricted range of $p\phi$ because of quadrant ambiguities and eventual division by zero. In practice, this problem is solved by a modified two-argument arctangent function well known as $\atantwo(a, b)$ (\autoref{fig:atan2-vs-atan}) \cite{Ukil2011}.
		   	
   		\subsection{Definition of the angular error} \label{sec:error-definition}
\begin{figure*}
	   		\centering
	   		\begin{subfigure}[t]{.49\textwidth}
	   			\centering
	   			\includegraphics[width=\columnwidth]{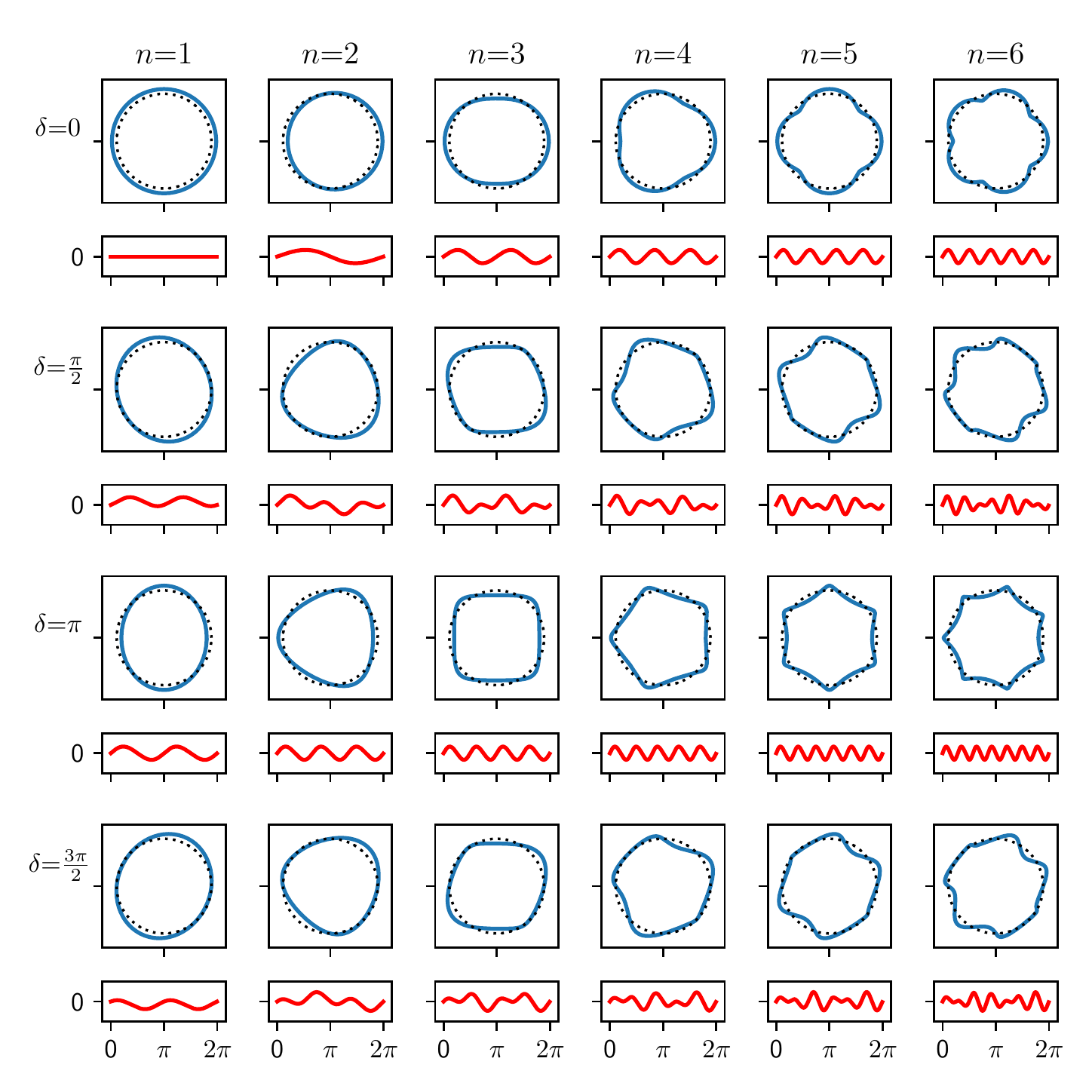}
	   			\caption{$A_n=B_n=0.1$}
	   			\label{fig:lissajous-overview-a}
	   		\end{subfigure}\hfill
	   		\begin{subfigure}[t]{.49\textwidth}
	   			\centering
	   			\includegraphics[width=\columnwidth]{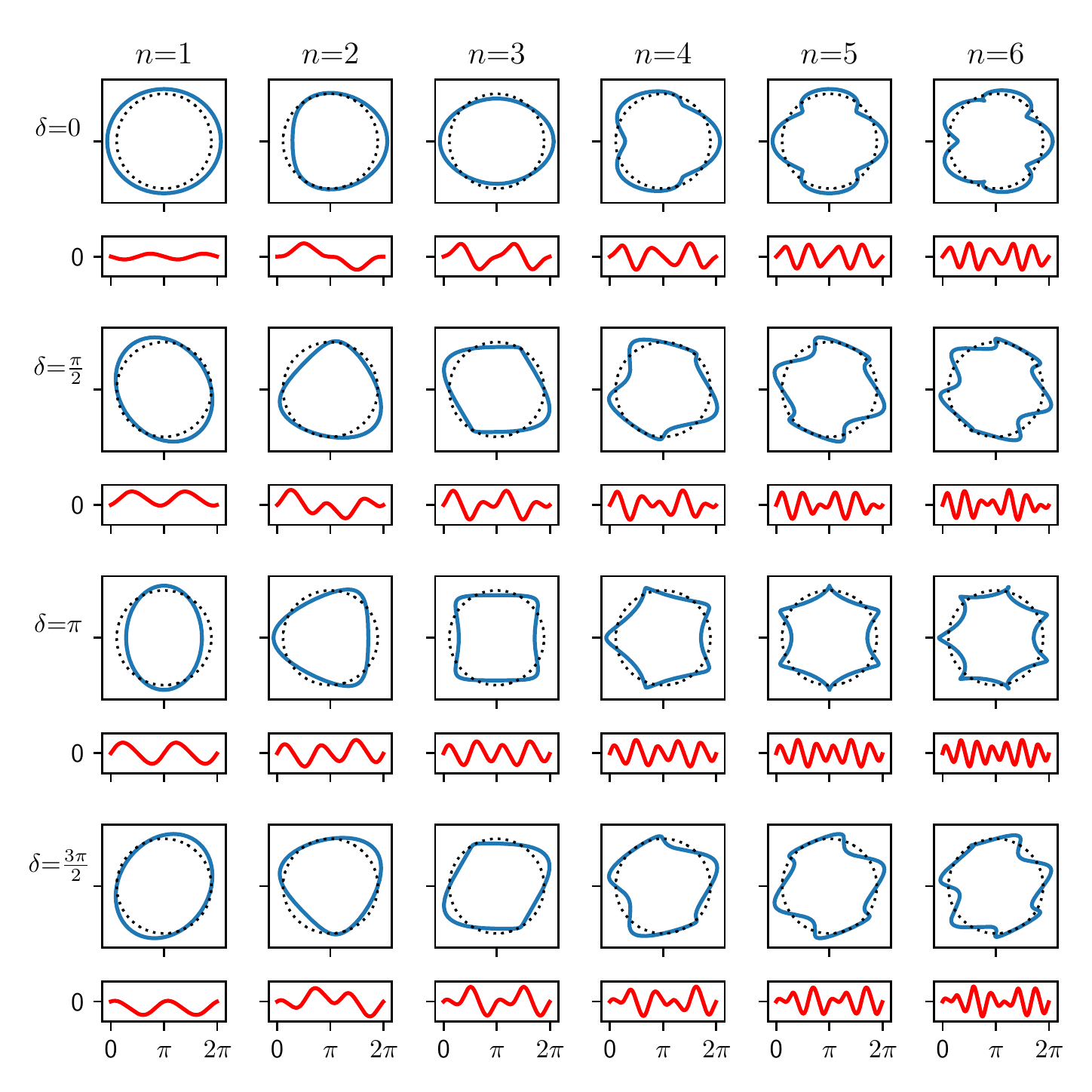}
	   			\caption{$A_n=0.1, B_n=0.2$}
	   			\label{fig:lissajous-overview-b}
	   		\end{subfigure}
	   		\caption{Lissajous figure examples (square) and corresponding angular error curves (below) for various signal distortions. The signals contain ideal main harmonics (unit amplitude) with one superimposed disturbance harmonic.  in $a$ and $b$, characterized by the harmonic order $n$, amplitudes $A_n$, $B_n$ and the phase difference $\delta_n$. Note that a superimposed 1\textsuperscript{st} harmonic is technically equivalent to an amplitude mismatch or phase error of the main harmonic.}
	   		\label{fig:lissajous-overview}
	   	\end{figure*}

	   		The encoder signals $a$ and $b$ form a vector $z$ in the complex plane. This vector results from the superposition of the main harmonic $z_p$, the offset $z_0$ and the sum of disturbing harmonics $z_n$. When varying the angle $\phi_p$ (\autoref{fig:vectors-complex-plane}), the vector $z$ consequently traces a \emph{curve}---also referred to as a \emph{Lissajous} figure. In case of ideal signals, the curve coincides with the unit circle traced by $e^{i\phi_p}$. When additional disturbance harmonics are present, the angle $\phi_z$ of the signal vector $z$ deviates from the electrical angle $\phi_p$.
	   		This deviation is quantified with the \emph{electrical} angular error $\Delta\phi_p$ which has a visible impact on the Lissajous figure. As an example, \autoref{fig:lissajous-overview} shows various patterns for a single disturbance harmonic.
		   	
\begin{figure}
	   		\centering
	   		\includegraphics[width=\columnwidth]{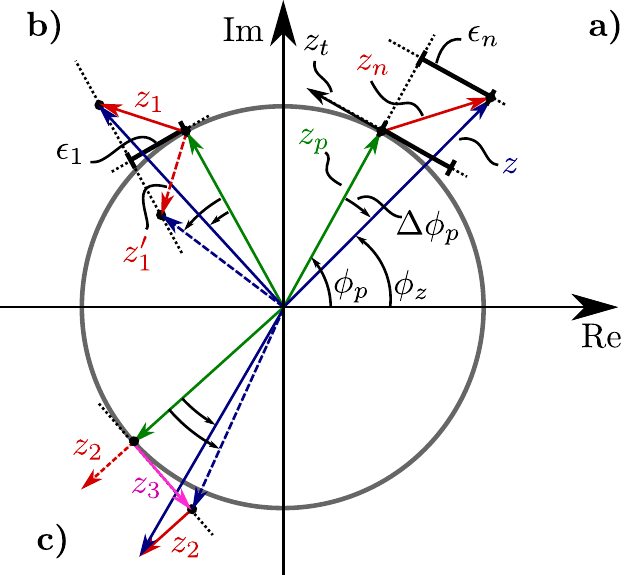}
	   		\caption{Geometric approximation $\epsilon_n$ of the angular error $\Delta\phi_p$ in the complex plane by projection of the disturbing harmonic $z_n$ onto the tangent $z_t$ of the unit circle which is traced by the ideal encoder signal $z_p$ (a). It is geometrically apparent that this first-order approximation does not include higher-order terms, because two disturbances $z_1$ and $z'_1$ with different angular errors can both lead to the same approximated error $\epsilon_1$ depending on the phase of the disturbance (b). An interaction effect between several disturbances exists, as $z_2$ alone does not lead to an angular error for the specific $\phi$ shown in the diagram. When combined with $z_3$, the angular error decreases compared to the effect of $z_3$ alone (c).}
	   		\label{fig:geometric-method}
	   	\end{figure}

		   	The electrical angular error be geometrically defined in the complex plane as the angle difference between $\phi_z$ and $\phi_p$ (\autoref{fig:geometric-method}a). Therefore
		   	\begin{align}
		   		\Delta\phi_p &= \phi_z - \phi_p\\
	   		\intertext{which leads to}
		   		\Delta\phi_p &=\atan(\frac{a(\phi)}{b(\phi)}) - p\phi\,.
		   		\label{eq:electrical-error}
		   	\end{align}
	   		
		   	Note that this formulation slightly differs from the geometrical definition, since the discontinuities of the arctangent function lead to discontinuities in the angular error curve, if $\phi\geq\frac{\pi}{2}$ or $\phi\leq-\frac{\pi}{2}$. The widely used two-argument arctangent ($\atantwo$) function does not fix this problem, it only reduces the number of discontinuities by half (\autoref{fig:atan2-vs-atan}). A practicable solution in numerical analyses is the use of a so-called \emph{phase unwrap} function, which is implemented in many scientific software tools.
		   	
 			With respect to Eq. \eqref{eq:mech-electr-angle}, the \emph{mechanical} angular error is defined as
		   	\begin{equation}
		   	\Delta\phi = \frac{1}{p}  \Delta\phi_p.
		   	\end{equation}
	   		Under the same conditions, the mechanical angular error therefore generally decreases with increasing periodicity $p>1$. However, the absolute angle can no longer be determined unambiguously without further measures. 
	   		
	   		When assessing the performance of an encoder in practice, the Fourier transform $\mathcal{F}$ of the angular error $\Delta \phi$ is often of interest, e.g. because higher order harmonics contribute more significantly to the error in rotor speed calculations. In the following, we use the notation $H_n$ for the amplitude of the harmonic of order $n$ in the frequency domain, disregarding the phase. Therefore, the natural unit of $H_n$ is radians. It can be obtained numerically by discrete Fourier transform of the angular error $\left | \mathcal{F}\left\{ \Delta \phi \right\}[n] \right |$ with appropriate normalization depending on the number of samples.
		   	
   		\subsection{Correction of the main signal harmonic}
			In order to reduce the angular error, basic transformations are commonly applied to the sensor output signals---often in the form of offset, amplitude and possibly orthogonality corrections which can be achieved by simple linear transformation of the signals (\autoref{eq:compensation}).
			
			Throughout the literature, numerous other methods are described, e.g. online correction methods based on an observer \cite{Albrecht2017}. Such methods are also able to compensate fluctuating error influences over lifetime, but will not be part of our analysis.
			 
			For the derivation of a simple compensation method, we consider two encoder signals $a$ and $b$ according to Eq. \eqref{eq:signals-fourier-definition}, but without additional disturbance harmonics for simplicity. The phase reference is arbitrarily chosen as the cosine channel $b$ and we define $\delta_p =\psi_p-\theta_p$ as the phase mismatch which yields
			\begin{align}
				\begin{split}
				a(\phi) &= A_p \sin(p\phi + \delta_p) + A_0,\\
				b(\phi) &= B_p \cos(p\phi) + B_0
				\end{split}
			\end{align}
			as signals. In practice, other disturbances might also be present (Eq. \eqref{eq:signals-fourier-definition}) which are also affected by the following compensation.
			
			Amplitudes and offsets are trivially corrected by
			\begin{equation}
				\begin{alignedat}{2}
					a' & = \frac{a - A_0}{A_p} &&= \sin(p\phi + \delta_p),\\
					b' &= \frac{b - B_0}{B_p} &&= \cos(p\phi).
				\end{alignedat}
			\end{equation}
			To ensure orthogonality of the signals $a$ and $b$ the phase mismatch can also be corrected.
			From Eq.\,\eqref{eq:trig-ortho} follows
			\begin{equation}
				a' = \sin(p\phi + \delta_p) = \sin(p\phi)\cos(\delta_p) + \cos(p\phi)\sin(\delta_p)
			\end{equation}
			and therefore
			\begin{align}
				\label{eq:compensation}
				\begin{split}
				a'' &= \sin(p\phi) = \frac{a' - \cos(p\phi)\sin(\delta_p)}{\cos(\delta_p)}\\
				&= \frac{\frac{a-A_0}{A_p} - \frac{b-B_0}{B_p} \sin(\delta_p)}{\cos(\delta_p)}\\
				&= \frac{(a-A_0) - (b-B_0) \frac{A_p}{B_p} \sin(\delta_p)}{A_p\cos(\delta_p)}.
				\end{split}
			\end{align}
	$a''$ and $b'$ then contain ideal main harmonics of unit amplitude that trace a unit circle in the complex plane (\autoref{fig:vectors-complex-plane}). As a prerequisite for the further sections we assume ideal main harmonics---however, our method can also be applied to determine the angular error caused by offset, amplitude mismatch and orthogonality of the main harmonic (\autoref{sec:special-cases}, \autoref{tab:table-special-cases}).
	
   	\section{Approximation of the angular error}
	   	Unfortunately, no Fourier transform of the arctangent function exists. In order to analyze the harmonic decomposition of the angular error, other methods have to be used instead. First of all, we present an intuitive geometric approximation, which is based on projections of the encoder signals in the complex plane. The result can be generalized to higher-order approximations by the use of a Taylor series expansion in terms of the harmonic distortion amplitudes.
	   	
	   	Initially, we assume ideal main harmonic signals without any offset, amplitude mismatch or orthogonality error as defined in Eq.~\eqref{eq:signal-assumptions}. Additionally, we assume that the amplitude of disturbances in the signal is much smaller than the usable main harmonic encoder signals, so that $A_n \ll A_p$ and $B_n \ll B_p$ for all harmonics $n$. This is appropriate for virtually all real-world applications since failing to meet this requirement results in significant angular error. Note that even if the encoder signals exhibit no
	   	amplitude mismatch, but are not normalized to unit amplitude, normalization is required (\autoref{sec:normalization}) because disturbances act proportionally to the amplitude of the main harmonic (implicit normalization occurs during evaluation of the arctangent function argument).
	   	
	   	\subsection{Geometric projection method} \label{sec:geometric-method}
 		In reference to \autoref{fig:geometric-method}a, consider an ideal main harmonic $z_p$ with one imposed disturbance $z_n$. As defined in Eq.~\eqref{eq:signal-definition}, both vectors rotate with different speeds corresponding to their harmonic order. \autoref{fig:geometric-method} therefore shows a snapshot in time (or rather: angle $\phi$) of the signals, wherein $z_p$ is assumed to be normalized and its trace coincides with the unit circle. 
		
		If the magnitude of $z_n$ is much smaller than $|z_p|=1$, the angular error $\Delta\phi_p$ is approximately equal to the projection of $z_n$ onto the tangent of the unit circle at $z_p$---recall the definition of an angle as the length of a segment on the unit circle. We denote this approximation $\epsilon_n$ in units of radians, which can be calculated for arbitrary amplitudes $A_n$, $B_n$ and phases $\theta_n$, $\psi_n$.
		
		By the use of Euler's formula, the definition of $z_n$ can be rewritten as
		  	\begin{equation}
		   		z_n = \frac{1}{2}\left(A_n e^{i(n\phi+\theta_n)}-A_n e^{-i(n\phi+\theta_n)}+B_n e^{i(n\phi+\psi_n)}-B_n e^{-i(n\phi+\psi_n)} \right).
		   	\end{equation}
		   	Correspondingly, the unit tangent $\hat{z_t}$ is given by
		   	\begin{equation}
		   		\hat{z_t} = ie^{ip\phi}.
		   	\end{equation}
		   	The projection onto the unit tangent $\hat{z_t}$ can be described as the multiplication of $z_n$ with the complex conjugate of $\hat{z_t}$ and then taking the real part
		   	\begin{align}
		   		\label{eq:geometric-result}
		   		\epsilon_n &= \Re\left\{\hat{\overline{z_t}} z_n\right\},\\
			   	\intertext{which yields}
		   		\begin{split}
				   	\epsilon_n &= \frac{1}{2}\biggl( A_n\sin((n-p)\phi+\theta_n)+B_n\sin((n-p)\phi+\psi_n)\\
				   	&\qquad+A_n\sin((n+p)\phi+\theta_n)-B_n\sin((n+p)\phi+\psi_n)\biggr).
			   	\end{split}
		   	\end{align}
		   	It is important to note that the angular error introduced by $z_n$ thus consists of harmonics of order $n - p$ and $n + p$, respectively, depending on the amplitudes $A_n$ and $B_n$ as well as the phases $\theta_n$ and $\psi_n$ of the disturbance. This becomes even more apparent by orthogonal decomposition:
		   	\begin{equation}
		   		\begin{split}
					\epsilon_n &= \frac12 \biggl( (A_n \sin\theta_n+B_n\sin\psi_n)\cos((n-p)\phi)\\
				   	&\qquad +(A_n \cos\theta_n+B_n\cos\psi_n)\sin((n-p)\phi)\\
				   	&\qquad +(A_n \sin\theta_n-B_n\sin\psi_n)\cos((n+p)\phi)\\
				   	&\qquad +(A_n \cos\theta_n-B_n\cos\psi_n)\sin((n+p)\phi)\biggr).
		   		\end{split}
		   	\end{equation}
		   	The amplitude of the angular error harmonics is therefore
			\begin{equation}
				H_{n-p} = \frac{1}{2}\sqrt{A_n^2+2A_nB_n\cos\delta_n+B_n^2}
			\end{equation}
		   	for the order $n-p$ of the angular error, and
		   	\begin{equation}
		   		H_{n+p} = \frac{1}{2}\sqrt{A_n^2-2A_nB_n\cos\delta_n+B_n^2}
		   	\end{equation}
		   	for the order $n+p$, where
		   	\begin{equation}
		   		\delta_n = \psi_n-\theta_n \,.
		   	\end{equation}

		   	Some interesting special cases arise which allow further simplification of the amplitudes. In case of orthogonal harmonics so that $\delta_n=0$
		   	\begin{equation}
			   	\begin{split}
			   		H_{n-p} &\underset{\delta_n=0}{=} \frac{A_n+B_n}{2}\\
			   		H_{n+p} &\underset{\delta_n=0}{=} \frac{|A_n-B_n|}{2}.
			   	\end{split}
		   	\end{equation}
		   	In case the amplitudes of the harmonics of order $n$ are equal so that $A_n = B_n$, but exhibit an orthogonality error $\delta_n$
		   	\begin{equation}
			   	\begin{split}
				   	H_{n-p} &\underset{A_n=B_n}{=} \frac{A_n}{\sqrt 2}\sqrt{1+\cos\delta_n}\\
				   	H_{n+p} &\underset{A_n=B_n}{=} \frac{A_n}{\sqrt 2}\sqrt{1-\cos\delta_n}.
			   	\end{split}
		   	\end{equation}
		   	If several harmonics $n, m, ...$ are present in the encoder signals $a$ and $b$, the approximated angular errors $\epsilon_n, \epsilon_m, ...$ can be superimposed by assuming independence of the effects of the disturbing harmonics. This is valid as a first-order approximation, though an interaction effect between harmonics exists in second order.
   	
		   	\autoref{fig:geometric-method}c shows an example for the interaction between two disturbing harmonics $z_2$ and $z_3$. When considering $z_2$ in relation to $z_p$, $z_2$ does not introduce any angular error at the current rotation position. On the other hand $z_3$ introduces an angular error when considered independently, though the error is reduced when regarding the superposition $z_2 + z_3$.
		   	
		   	Further, \autoref{fig:geometric-method}b illustrates why this first-order approximation is only valid for small disturbances. $z'_1$  has the same magnitude as $z_1$ but is mirrored on the tangent of $z_p$. The projection on the tangent yields the same approximation $\epsilon_1$ to the angular error in both cases, however, the exact angular error caused by $z'_1$ is larger.
		   	
		   	As we show in the next section through the derivation of a second-order approximation, these effects are often negligible in real-world applications where the amplitude of the disturbances is sufficiently small compared to the main harmonic. 
   	
   		\subsection{Taylor series approximation}
   		\label{sec:taylor-approx}
   			A more powerful approach to approximate the angular error can be obtained by multivariate Taylor-Maclaurin series expansion.
		   	As defined in Eq.\,\eqref{eq:electrical-error}, the \emph{electrical} angular error $\Delta\phi_p$ results from combining the signals $a$ and $b$ with the sum of disturbing harmonics
		   	\begin{align} \label{eq:error-definition}
		   		\begin{split}
			   	\Delta\phi_p &= \atan(\frac{\sin(p\phi)+\sum\limits_{n}A_n\sin(n\phi+\theta_n)}{\cos(p\phi)+\sum\limits_{n}B_n\cos(n\phi+\psi_n)})-p\phi\\
			   	&=f(\mathbf{A},\mathbf{B}) \,.
			   	\end{split}
		   	\end{align}
		   	We denote this function $f$ for convenience, where $\mathbf{A}$ and $\mathbf{B}$ are the amplitudes of the disturbances in $a$ and $b$ respectively, expressed as vectors.
		   	
		   	Under the same assumption as made for the geometric approximation---that the amplitudes of disturbances are small compared to the main harmonic---$f$ can be approximated by the multivariate Maclaurin series of order $k$ with remainder $R_k$ (Eq.\,\eqref{eq:multivariate-maclaurin}).
		   	By simplification of the first and second order we obtain
			\begin{align}
				\label{eq:taylor-error}
				\begin{split}
				   	f &= \overbrace{\cancel{\atan(\frac{\sin(p\phi)}{\cos(p\phi)}) - p\phi}}^{\overset{\mathrm{def}}{=} 0}\\
			   	   	&+\frac{1}{2}\sum\limits_{n}\biggl(A_n\sin((n-p)\phi+\theta_n)+B_n\sin((n-p)\phi+\psi_n)\\
				   	&\qquad+A_n\sin((n+p)\phi+\theta_n)-B_n\sin((n+p)\phi+\psi_n)\biggr)\\
			   	   	&+\frac{1}{2}\sum\limits_{n}\sum\limits_{m}\biggl(-A_n A_m\sin(2p\phi)\sin(n\phi+\theta_n)\sin(m\phi+\theta_m)\\
				   	&\qquad\qquad-2 A_n B_m\cos(2p\phi)\sin(n\phi+\theta_n)\cos(m\phi+\psi_m)\\
				   	&\qquad\qquad+B_n B_m\sin(2p\phi)\cos(n\phi+\psi_n)\cos(m\phi+\psi_m)\biggr)\\
				   	&+R_2(\mathbf{A},\mathbf{B}).
			   	\end{split}
			\end{align}
			The above Taylor expansion is composed of the terms\footnote{Note that all harmonic orders ($p, n, m$) refer to one full period of the reference angle and are therefore \emph{mechanical} harmonics.}:
			\begin{flalign}
				\begin{split}
				&\left.\begin{aligned}
				f =& \;\;T^0\\
				\end{aligned}\quad\right\} \overset{\mathrm{def}}{=} 0\\
				&\left.\begin{aligned}
				&\quad+T^1
				\end{aligned}\right\}\begin{array}{l}
				\text{1\textsuperscript{st} order approximation}\\
				\text{Harmonics } p \pm n
				\end{array}\\
				&\left.\begin{aligned}
				&\quad+T^2
				\end{aligned}\right\}\begin{array}{l}
				\text{2\textsuperscript{nd} order approximation}\\
				\text{Harmonics } 2p\pm n\pm m
				\end{array}\\
				&\left.\begin{aligned}
				& \quad+ R_2(\mathbf{A}, \mathbf{B})
				\end{aligned}\right\}\text{Remainder}
				\end{split}
			\end{flalign}
						
		   	Note that the first-order term $T^1$ matches the geometrically obtained result from section \ref{sec:geometric-method} (Eq. \eqref{eq:geometric-result}), yielding the possible harmonic orders $p + n$ and $p - n$ of the angular error. Both, interaction effects and an additional error (\autoref{fig:geometric-method}) can be explained by the second-order term $T^2$ of the Taylor approximation. 
		   	When $n\neq m$ in the summation, the term describes an interaction effect of orders $n$ and $m$, when $n=m$ it describes a second-order effect of the same harmonic. Consequently all possible combinations of $2p \pm n \pm m$ may occur.
		   		   	
		   	The zeroth-order of the Taylor series, represented by the term $T^0$, is a tricky case. As discussed in section \ref{sec:error-definition}, in numerical analyses the extended two-argument arctangent function is used to correct for quadrant ambiguities and the result is commonly \emph{unwrapped}, i.e. discontinuities are removed numerically. Therefore, the result of the modified arctangent operation follows a similar linear shape as the reference $p\phi$, but with an added angular error. Since $T^0$ contains only the ideal sine and cosine terms in the arctangent argument, any non-zero term when subtracting $p\phi$ is only due to the discontinuities of the arctangent function and ambiguities between electrical and mechanical angles when $p>1$.  Therefore, we \emph{define} it to be zero in order to match the numerical process of unwrapping the $\atantwo$ function.
		   	
		   	The amplitudes of angular error harmonics in the second-order term $T^2$ are decreased significantly compared to the first-order term $T^1$ since we assume $A_n \ll A_p = 1$ and $B_n \ll B_p = 1$ for every disturbance in the Fourier series. Therefore, the products $A_nA_m$, $A_nB_m$ and $B_nB_m$ always yield smaller values compared to the first-order approximation. It is possible to extend \autoref{eq:taylor-error} to include arbitrary higher orders $k>2$ based on Equations \ref{eq:multivariate-maclaurin} and \ref{eq:partial-d-general-solution}, however this yields diminishing returns in practice. For many applications, even a first-order approximation is sufficient.
   	
   		\subsection{Error bounds}
		   	\begin{figure}
		   		\centering
		   		\includegraphics[width=\columnwidth]{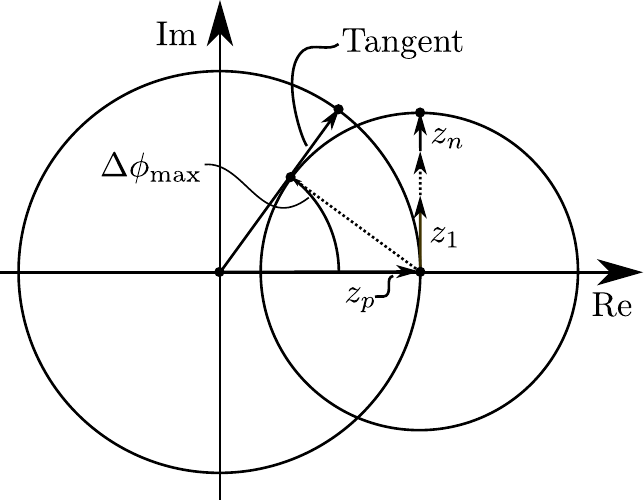}
		   		\caption{Geometric bound for the angular error due to the disturbance harmonics $z_1$, ..., $z_n$ superimposed onto the ideal main harmonic $z_p$ of an angle encoder in the complex plane (Eq. \eqref{eq:signal-definition}). By assuming a worst-case superposition of harmonics, the maximum angular error $\Delta\phi_{max}$ is given by the angle of the tangent to the disturbance Lissajous figure (circular due to worst-case superposition). The actual maximum of the angular error may be lower due to the phase relations of the disturbance harmonics.}
		   		\label{fig:geometric-error-bound}
		   	\end{figure} 
		   	It is useful to determine the maximum approximation error, which can be achieved by finding an upper bound for $R_k(\mathbf{A},\mathbf{B})$. For comparison, we will first derive a bound for the maximum angular error geometrically, given the sum of disturbance harmonics.
		   	
		   	In reference to \autoref{fig:geometric-error-bound}, let $\mathcal{A}$ be the sum of the maximum magnitudes of the disturbing harmonics
		   	\begin{equation} \label{eq:mean-squared-amp}
		   		\mathcal{A} = \sum\limits_{n} \sqrt{A_n^2+B_n^2} \enspace <1.
		   	\end{equation}
		   	The angular error is then bounded by
		   	\begin{equation}
		   		\label{eq:geometric-upper-bound}
		   		\Delta\phi_p \leq \atan \left(\frac{\mathcal{A}\sqrt{1-\mathcal{A}^2}}{1-\mathcal{A}^2}\right)
		   	\end{equation}
		   	which is achieved by finding the tangent from the origin to the circle with radius $\mathcal{A}$ centered around $z_c=1+0i$. The above bound is tight in the worst-case when the phases of the disturbance harmonics align in such a way that causes the maximum angular error in relation to $z_p$.
		   	
		   	If $\mathcal{A}<\frac{1}{2}$, which is a reasonable assumption for most encoders, then
		   	\begin{equation}
		   		\Delta\phi_p = \leq \frac{\pi}{3}\mathcal{A}
		   	\end{equation}
		   	is a good approximation that avoids the arctangent function and is often handy for rule-of-thumb assessments ($\frac{\pi}{3} = \ang{60}$).
		   	
		   	A bound for the residual $R_k$ of the Taylor series can also be found for arbitrary
		   	approximation orders $k$. Similarly to Eq.\,\eqref{eq:mean-squared-amp}, let 
		   	\begin{equation}
			   	\tilde{\mathcal{A}} = \sum_{n}\left(A_n + B_n \right) = \norm{\mathbf{A}}_1 + \norm{\mathbf{B}}_1 
		   	\end{equation}
		    then the remainder of the $k^\text{th}$-order Taylor series is bounded by
		   	\begin{equation}
			   	\left | R_k(\mathbf{A}, \mathbf{B}) \right | \le 
			   	-\log(1-\tilde{\mathcal{A}}) - \sum\limits_{q = 1}^{k} \frac{1}{q}{\tilde{\mathcal{A}}^q} \,.
		   	\end{equation}
		 	Refer to \ref{sec:appendix-taylor} for a derivation of this error bound.

   	\section{Numerical example} \label{sec:numerical-validation}
  		\begin{figure}
	   		\centering
	   		\includegraphics[width=\columnwidth]{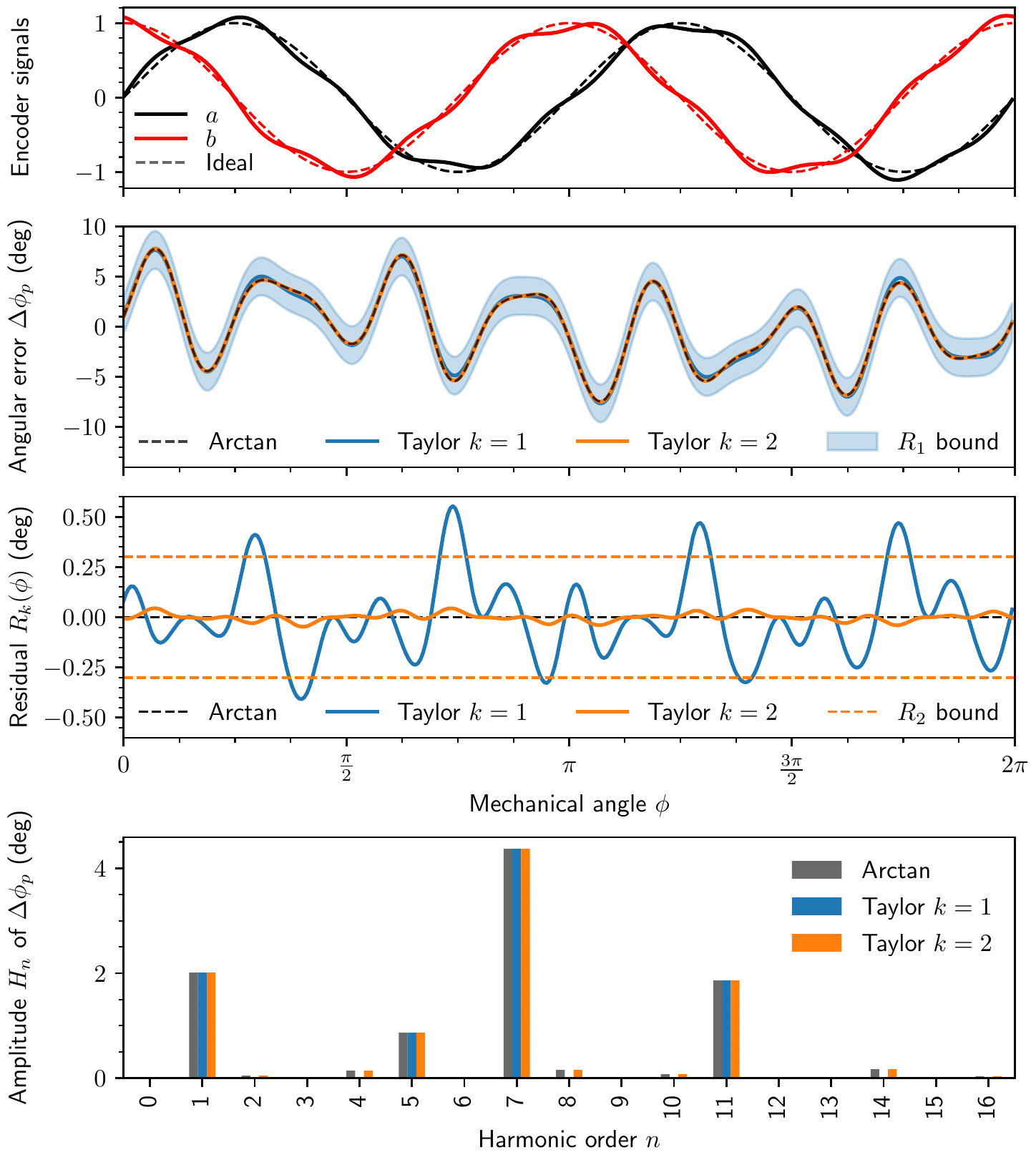}
	   		\caption{Comparison of the electrical angular error $\Delta\phi_p$ as calculated numerically with the arctangent function and the Taylor series approximation of first and second order for the encoder signals $a$ and $b$ (Eq.\,\eqref{eq:numerical-example}). The bound of the residual error of the approximation is given for the Taylor series residual with $k=1$ (shaded blue area) and $k=2$ (orange horizontal lines). The bottom plot shows the decomposition of the angular error into harmonic amplitudes, as estimated by the Taylor series and calculated by discrete Fourier transform for the numerical arctangent function.}
	   		\label{fig:taylor-comparison}
   		\end{figure} 
   	
   		In order to validate the presented approximation method from section \ref{sec:taylor-approx}, the following signal is used as an example for comparison between the arctangent function and the results from Eq.\,\eqref{eq:taylor-error}:
		\begin{align}
			\begin{split}
			\label{eq:numerical-example}
			a(\phi) &= \sin(2\phi) + 0.05 \sin(3\phi + \frac{\pi}{8}) + 	0.075\sin(9\phi),\\
			b(\phi) &= \cos(2\phi) +  0.02 \cos(3\phi + \frac{\pi}{7}) + 	0.09\cos(9\phi+\frac{\pi}{4}).
			\end{split}
		\end{align}
		This example contains a main signal harmonic with the order $p=2$ and two added disturbing harmonics of the order 3 and 9, including an amplitude mismatch and orthogonality error of the harmonics.
		
		The result of the Taylor approximation of order $k=1$ and $k=2$ is presented in \autoref{fig:taylor-comparison}. Given a large angular error caused by the harmonics (max $\sim\ang{8}$), the Taylor series approximation yields excellent results. The maximum value of the residual is smaller than $\ang{0.6}$ for the first-order approximation and smaller than $\ang{0.05}$ for the second order. When assessing the harmonic decomposition of the angular error, all dominant orders (1, 5, 7, 11) are excellently approximated by the first-order approximation. In second order, minor harmonics with small amplitude (2, 4, 8, 10, 14, 16) are also approximated correctly. 
   	
   	\section{Common distortions of encoder signals}
	\label{sec:special-cases}
   		To avoid ambiguities and to have a clear distinction between ideal main signal harmonic and higher-order disturbing harmonics, we assumed amplitude-, offset- and phase-corrected signals. Thus, the influence of such mismatches were excluded from the derivations in chapters \ref{sec:geometric-method} and \ref{sec:taylor-approx}. However, Eq.\,\eqref{eq:taylor-error} also allows to predict the composition of the angular error in case these commonly encountered mismatches are present in the signal domain. This is easily possible by treating deviations of the main harmonic from the ideal signal as equivalent disturbance harmonics of order $n=0$ and $n=p$.
   		
   		In \autoref{tab:table-special-cases} the harmonic amplitudes of the angular error are listed for all possible mismatch combinations, calculated with the second-order Taylor series from Eq. \eqref{eq:taylor-error}.
   		
   		For the approximation of the angular error caused by a DC offset of the signals $a$ and $b$, we choose $n=0$ as the order of the disturbance harmonics and define the phase of the imaginary part $\theta_n=\pi/2$ to avoid cancellation of the sine term in $a$ while the phase of the real part $b$ is simply $\psi_n=0$. Negative amplitudes can be allowed for convenience.
   		
   		A mismatch of the sine and cosine amplitudes of the main signal can be achieved by setting $n=p$. The disturbance harmonics amplitudes $A_n$ and $B_n$ therefore represent the deviation of the main signal from the ideal case of unit amplitudes.
   		
   		To calculate the influence of a phase mismatch, we again make use of the trigonometric identity in Eq.\,\eqref{eq:trig-ortho} and choose the cosine channel as the reference:
   		\begin{equation}
   			\label{eq:orthogonality-error-decomposition}
	   			\sin(p\phi+\delta_p) = \sin(p\phi)+2\sin(\frac{\delta_p}{2})\sin(p\phi+\frac{\delta_p+\pi}{2}).
   		\end{equation}
		As this result shows we have to choose $n=p$, $A_n=2\sin(\frac{\delta_p}{2})$ and $\theta_n=\frac{\delta_p+\pi}{2}$ to represent the phase mismatch in the signal domain. 
   		
   		\subsection{Normalization}
   		\label{sec:normalization}
   		Consider the signals
   		\begin{align}
   			\label{eq:example-offset-amplitude-phase}
   			\begin{split}
   				a &= A_p\sin(p\phi+\theta_p) + A'_0\\
   				b &= B_p\cos(p\phi) + B'_0
   			\end{split}
   		\end{align}
   		Because disturbances affect the angular error inversely proportional to the amplitude of the main harmonic, it is necessary to normalize the signals $a$ and $b$ by scaling them (and therefore the Lissajous figure) both with the same factor. Note that this has no effect on the calculated angle, because the scaling term cancels out in the fraction of the arctangent argument (Eq. \eqref{eq:atan}). Because of the amplitude mismatch present in the signals, it is impossible to scale both of them exactly to unit amplitude, e.g. in the case $A_p = 2\,\text{Volts}$, $B_p = 2.5\,\text{Volts}$. In order to achieve an even scaling for both sine and cosine channel, we propose that the mean of the amplitudes may be chosen as the scaling factor $g$. Following the example in Eq. \eqref{eq:example-offset-amplitude-phase}
   		\begin{equation}
   			g = \frac{A_p + B_p}{2} \,.
   		\end{equation}
   		The equivalent \enquote{offset harmonic} with $n=0$ therefore evaluates to
   		\begin{equation}
   			z_0 = i \underbrace{\frac{A'_0}{g}}_{=A_0} \underbrace{\vphantom{\frac{A_0}{g}} \sin(0\phi + \frac{\pi}{2})}_{=1} + \underbrace{\frac{B'_0}{g}}_{=B_0} \underbrace{\vphantom{\frac{A_0}{g}}\cos(0\phi)}_{=1}
   		\end{equation}
   		where $A_0$ and $B_0$ are the offsets that can be found in \autoref{tab:table-special-cases}.
   		The amplitude mismatch is analogously expressed as
   		\begin{equation}
   			z_n = i \underbrace{\left(\frac{A_p}{g}-1\right)}_{A_n} \sin(p\phi) + \underbrace{\left(\frac{B_p}{g}-1\right)}_{B_n} \cos(p\phi) \,.
   		\end{equation}
   		by taking the difference to unit amplitude after normalization as the amplitude of the disturbance harmonic. Note that the normalization applies to possibly present higher-order harmonics as well. The expression of the phase mismatch (Eq. \eqref{eq:orthogonality-error-decomposition}) as an equivalent harmonic requires no normalization of amplitude.
   		
   		The effect of phase and amplitude mismatch is considered as two separate disturbance harmonics, although they both lead to an effective disturbance of the same order ($2p$). The interaction effect of phase and amplitude mismatch (as well as all other interactions) is still considered in the second-order terms of the Taylor series. This approach avoids the complexities in notation involved with the aggregation of both mismatches into a single harmonic.
   		
   		\begin{table*}
   			\centering
   			\resizebox{1\textwidth}{!}{\begin{tabular}{L{3.2cm}|L{4.5cm}|L{3.2cm}|l}
   				\multicolumn{2}{c|}{\tab[0.3cm]\Large\textbf{Signal domain}} & \multicolumn{2}{l}{\tab[2.25cm]\Large\textbf{Error domain}}\\
   				\textbf{Mismatch} & \textbf{Equivalent Disturbance} & \textbf{Orders} & \textbf{Harmonic amplitudes}\\
   				\hline
   				Offset & $B_0+iA_0$ & \makecell[l]{$T^1: p$\\$T^2: 2p$} & \parbox{2cm}{\begin{align*}
   						H_p&=\sqrt{A_0^2+B_0^2}\\
   						H_{2p}&=\sqrt{\frac{A_0^4}{4} + \frac{A_0^2B_0^2}{2} + \frac{B_0^4}{4}}
   				\end{align*}}\\
   				Amplitude & \makecell[l]{$B_n\cos(p\phi)$\\$+iA_n\sin(p\phi)$} & \makecell[l]{$T^1: 2p$\\$T^2: 4p$} & \parbox{2cm}{\begin{align*}
   						H_{2p} &=\frac12(A_n-B_n) + \frac14 (B_n^2 - A_n^2)\\
   						H_{4p} &= \frac18 (A_n - B_n)^2
   				\end{align*}}\\[6pt]
   				Phase & $i2\sin(\frac{\delta_p}{2})\sin(p\phi+\frac{\delta_p+\pi}{2})$ & \makecell[l]{$T^1: 0,2p$\\$T^2: 0,2p,4p$}& \parbox{2cm}{\begin{align*}
   						H_0&=\frac34\sin\delta_p-\frac18\sin(2\delta_p)\\
   						H_{2p}&= \sqrt{\frac14 \sin^2\delta_p + (\cos\delta_p-1)^2} \approx \frac12\delta_p\\
   						H_{4p}&=\frac14(1-\cos\delta_p)
   				\end{align*}}\\
   				\makecell[l]{Offset,\\amplitude}
   				&\makecell[l]{$B_0+B_n\cos(p\phi)$\\$+i(A_0+A_n\sin(p\phi))$} & \makecell[l]{$T^1: p,2p$\\$T^2: p,2p,3p,4p$} & \parbox{5cm}{\begin{align*}
   						H_p&=\frac12(B_n+A_n-2)\sqrt{B_0^2+A_0^2}\\
   						H_{2p}&=\sqrt{B_0^2A_0^2+\frac{1}{16}(2B_0^2+B_n^2-2B_n-2A_0^2-A_n^2-2A_n)^2}\\
   						H_{3p}&=\frac12(A_n-B_n)\sqrt{B_0^2+A_0^2}\\
   						H_{4p}&=\frac18(A_n-B_n)^2
   				\end{align*}}\\
   				\makecell[l]{Offset,\\phase} & \makecell[l]{$B_0$\\$+iA_0$\\$+i2\sin(\frac{\delta_p}{2})\sin(p\phi+\frac{\delta_p+\pi}{2})$} & \makecell[l]{$T^1: 0,p,2p$\\$T^2: 0,p,2p,3p,4p$} & \parbox{5cm}{\begin{align*}
   						H_0&=\frac34\sin\delta_p-\frac18\sin(2\delta_p)\\
   						H_p&=\frac{\sqrt{2}}{2}\sqrt{(5-3\cos\delta_p)(A_0^2+B_0^2)}\\
   						H_{2p}&=\frac12\sqrt{(\sin\delta_p-2A_0B_0)^2+(2\cos\delta_p-A_0^2+B_0^2-2)^2}\\
   						H_{3p}&=\frac{\sqrt{2}}{2}\sqrt{(1-\cos\delta_p)(A_0^2+B_0^2)}\\
   						H_{4p}&=\frac14(1-\cos\delta_p)
   				\end{align*}}\\
   				\makecell[l]{Amplitude,\\phase} & \makecell[l]{$B_n\cos(p\phi)$\\$+iA_n\sin(p\phi)$\\$+i2\sin(\frac{\delta_p}{2})\sin(p\phi+\frac{\delta_p+\pi}{2})$} & \makecell[l]{$T^1: 0,2p$\\$T^2: 0,2p,4p$} & \parbox{5cm}{\begin{align*}
   						H_0&=\frac14(3-A_n-B_n)\sin\delta_p-\frac18\sin(2\delta_p) \\
   						H_{2p}&=\frac14\sqrt{(A_n^2-B_n^2-4A_n+2B_n+2A_n\cos\delta_p-4\cos\delta_p+4)^2+4\sin^2\delta_p(B_n-1)^2}\\
   						H_{4p}&=\frac18\sqrt{
   							\begin{gathered}%
   								(\sin(2\delta_p)+2\sin\delta_p(A_n-B_n-1))^2+((A_n-B_n)^2\\
   								-2(A_n-B_n)+2\cos\delta_p(A_n-B_n-1)+\cos(2\delta_p)+1)^2
   							\end{gathered}%
   						}
   				\end{align*}}\\
   				\makecell[l]{Offset,\\amplitude,\\phase}& \makecell[l]{$B_0+B_n\cos(p\phi)$\\$+i(A_0+A_n\sin(p\phi))$\\$+i2\sin(\frac{\delta_p}{2})\sin(p\phi+\frac{\delta_p+\pi}{2})$} & \makecell[l]{$T^1: 0,p,2p$\\$T^2: 0,p,2p,3p,4p$} &\parbox{5cm}{\begin{align*}
   						H_0&=\frac14(3-A_n+B_n)\sin\delta_p-\frac18\sin(2\delta_p)\\ H_p&=\frac12\sqrt{(A_0^2+B_0^2)\left((A_n+B_n)^2-6(A_n+B_n)+2\cos\delta_p(A_n+B_n)-6\cos\delta_p+10\right)}\\
   						H_{2p}&=\frac12\sqrt{
   							\begin{aligned}%
   								(&\sin\delta_p(B_n-1)+2A_0B_0)^2\\
   								+&\Bigl(\frac12(A_n^2-B_n^2)+A_0^2-B_0^2-2A_n+B_n+\cos\delta_p(A_n-2)+2\Bigr)^2
   							\end{aligned}%
   						}\\
   						H_{3p}&=\frac12\sqrt{(A_0^2+B_0^2)\left((A_n-B_n)^2-2(A_n-B_n)+2\cos\delta_p(A_n-B_n-1)+2\right)}\\
   						H_{4p}&=\frac18\sqrt{%
   							\begin{aligned}%
   								(&\sin(2\delta_p)+2\sin\delta_p(A_n-B_n-1))^2\\
   								+&((A_n-B_n)^2 -2(A_n-B_n)+2\cos\delta_p(A_n-B_n-1)+\cos(2\delta_p)+1)^2%
   							\end{aligned}%
   						}%
   				\end{align*}}\\
   			\end{tabular}}
   			\caption{Common mismatches of the main harmonic of the encoder signals and resulting second-order approximation of the harmonic amplitudes of the electrical angular error in units of radians.  
   			Encoder signal offsets, amplitude mismatch and phase (orthogonality) errors are expressed as an equivalent disturbance harmonic of order $n$ equal to the periodicity $p$, so they can be analyzed with the Taylor series approximation developed in section \ref{sec:taylor-approx}. $T^1$ refers to the first-order Taylor terms, $T^2$ to the second-order terms. Corresponding harmonic orders of the angular error are expressed in dependence of the periodicity $p$ of the encoder.
   			}
   			\label{tab:table-special-cases}
   		\end{table*}
   		
   	\section{Conclusion}
    	In order to estimate the harmonic composition of the angular error due to disturbance harmonics in the encoder signals, we developed a two powerful methods to accurately approximate the angular error harmonics of the arctangent function: An intuitive geometric approach in the complex plane and a Taylor series approximation in terms of the distortion amplitudes. Both yield the same results in a first-order approximation. The geometric interpretation provides a vivid explanation of the source of angular error harmonics. On the other hand, the Taylor series can be extended to higher-order approximations, taking into account second-order effects and interactions between disturbance harmonics.
    	
    	Both approximation methods are in excellent agreement with the numerical examples (\autoref{sec:numerical-validation}). In many cases in practice, the first-order Taylor series  is sufficient, yielding possible harmonic orders of $n-p$ and $n+p$ for the angular error, where $n$ is the order of a disturbance harmonic and $p$ the periodicity of the encoder signals. 
    	A more accurate result can be achieved by also considering the second-order. The result shows no significant deviation between the numerically evaluated arctangent function and our approximation (Eq.\,\eqref{eq:taylor-error}), even in cases with large harmonic distortion amplitudes. Furthermore, we provide an upper bound for the residuals of the Taylor series and compare it with a geometric bound in the complex plane.

    	Additionally, offset-, amplitude- and orthogonality errors of the main harmonic of the encoder can also be analyzed with the obtained Taylor series. Our method is capable of predicting these distortions by representing them as equivalent disturbance harmonics. \autoref{tab:table-special-cases} gives a comprehensive overview about the resulting harmonics in the angular error.
    	
    	In general, our work contributes to a deeper understanding of the nonlinearity of angle encoders and enables thorough analysis of the \emph{cause} of observed harmonics in encoder designs. Further, it has applications in the future development of advanced error compensation algorithms and encoder models. For example, automated optimization of sensor geometry to minimize angular error harmonics can be simplified by replacing the arctangent calculation with our approximation in a virtual sensor design process.
    	 
	\section*{Funding}
	This research was funded by Robert Bosch GmbH -- Chassis Systems Control, Germany.

    \appendix
    \section{Partial derivatives of the arctan function}
	Throughout this paper, a number of trigonometric identities and derivatives are utilized. They can be found in the common literature, e.g. by \citet{Bronshtein2013}.
    
    The identity for orthogonal decomposition
    \begin{equation}
    	\label{eq:trig-ortho}
    	\sin \left(\alpha + \beta \right) = \sin\alpha \cos\beta + \cos\alpha\sin\beta
    \end{equation}
	is often needed. The uncommon identity 
    \begin{equation}
    	\left (\frac{\sin^2 \phi}{\cos^2 \phi} + 1 \right)\cos^k \phi = \left(\cos \phi\right)^{k-2}
    \end{equation}
    is particularly useful for the calculation of partial derivatives.
    
	    The derivatives required for the derivation of the Taylor series are quite unwieldy, therefore we recommend using a modern computer algebra system \cite{Meurer2017} together with trigonometry-specific simplification algorithms \cite{Fu2006} for calculations. Nonetheless, we provide an outline of the process \enquote{by hand} for verification.
	    
	    Let
	    \begin{equation}
		    \label{eq:def-f}
		    f = \atan\left(\frac{\sin p\phi + \sum\limits_n A_n\sin(n\phi + \theta_n)}{\cos p\phi + \sum\limits_n  B_n\cos(n\phi + \psi_n)}\right) - p\phi \,.
	    \end{equation}
	    Starting from the well-known trigonometric identities and derivatives of the arctangent function we obtain
	    \begin{align}
		    \begin{split}
			    \pdv{f}{A_m} &= \frac{\sin(m\phi + \theta_m)}{\left(1 + \left(\frac{\sin p\phi + \sum\limits_{n} A_n \sin(n\phi + \theta_n)}{\cos p\phi + \sum\limits_{n} B_n \cos(n\phi + \psi_n)}\right)^2\right)\left(\cos p\phi + \sum\limits_{n} B_n \cos(n\phi+\psi_n)\right)}.
		    \end{split}
		    \intertext{Evaluating the derivative at zero}
		    \begin{split}
			    \pdv{f}{A_m} &\bigg\rvert_{\mathbf{A} = \mathbf{B} = \mathbf{0}} = \frac{\sin{\left(m\phi + \theta_m \right)}}{\left(1 + \frac{\sin^{2}{\left(p\phi \right)}}{\cos^{2}{\left(p\phi \right)}}\right) \cos{\left(p\phi \right)}}
			    = \sin(m\phi + \theta) \cos (p\phi) \,.
		    \end{split}
  		\end{align}
  		Similarly,
		\begin{equation}
			\pdv{f}{B_m} = -\frac{\left(\sin p\phi +\sum\limits_{n} A_n \sin(n\phi + \theta_n)\right) \cos(m\phi + \psi_m)}{\left(1 + \left(\frac{\sin p\phi + \sum\limits_{n} A_n \sin(nx + \theta_n)}{\cos p\phi + \sum\limits_{n} B_n \cos(n\phi + \psi_n)}\right)^2\right) \left(\cos p\phi +\sum\limits_{n} B_n \cos(n\phi + \psi_n)\right)^2}%
		\end{equation}
		and
		\begin{equation}
			 \pdv{f}{B_m} \bigg\rvert_{\mathbf{A} = \mathbf{B} = \mathbf{0}} = -\frac{\sin p\phi \cos(m\phi + \psi_m)}{\left(1 + \frac{\sin^2 p\phi}{\cos^2 p\phi}\right)\cos^2 p\phi}
			 = -\sin p\phi \cos(m\phi + \psi_m) \,.%
		\end{equation}
  		\begin{align}
		    \intertext{The second-order derivatives evaluated at zero can be obtained in the same manner, yielding}
		    \begin{split}
			    \pdv{f}{A_n}{A_m} &\bigg\rvert_{\mathbf{A} = \mathbf{B} = \mathbf{0}} = -\sin(2p\phi) \sin(n\phi+\theta_n)  \sin(m\phi+\theta_m),
		    \end{split}\\
		    \begin{split}
			    \pdv{f}{B_n}{B_m} &\bigg\rvert_{\mathbf{A} = \mathbf{B} = \mathbf{0}} = \phantom{-}\sin(2p\phi) \cos(n\phi+\psi_n) \cos(m\phi+\psi_m), 
		    \end{split}\\
		    \begin{split}
			    \pdv{f}{A_n}{B_m} &\bigg\rvert_{\mathbf{A} = \mathbf{B} = \mathbf{0}} = -\cos(2p\phi) \sin(n\phi+\theta_n) \cos(m\phi+\psi_m)\,.
		    \end{split}
	    \end{align}
    
    \section{Multivariate Taylor series}
    \label{sec:appendix-taylor}
	We consider the multivariate Taylor series of order $k$ evaluated at zero, forming the multivariate Maclaurin series:
	    \begin{align}
	    	\label{eq:multivariate-maclaurin}
	    	f(\mathbf{A}, \mathbf{B}) = \sum\limits_{\abs{\alpha} + \abs{\beta} \le k} \frac{\partial^\alpha\partial^\beta f(\mathbf{0}, \mathbf{0})}{\alpha! \beta!} \mathbf{A}^\alpha \mathbf{B}^\beta + R_k(\mathbf{A}, \mathbf{B}).
	    \end{align}
	
		Foregoing a formal proof, given $f$ from Eq.\,\eqref{eq:def-f} we observe the following general solution for the partial derivative of $f$ evaluated at zero
		\begin{align}
			\label{eq:partial-d-general-solution}
			\begin{split}
				\partial^{\alpha}\partial^{\beta} &f(\mathbf{A} =\mathbf{0}, \mathbf{B} = \mathbf{0}) \\ &= (\abs{\alpha}+\abs{\beta}-1)! \sin((\abs{\beta}+\abs{\alpha})\phi + (\abs{\alpha}+2\abs{\beta}) \frac{\pi}{2}) \mathbf{a}^\alpha \mathbf{b}^\beta 
			\end{split}
		\end{align}
	
		using the dual multi-index notation
		\begin{align}
			\begin{split}
				\mathbf{A} &= (A_1, A_2, \dots, A_n)\\
				\mathbf{B} &= (B_1, B_2, \dots, B_n)
			\end{split}\\
			\begin{split}
				\alpha &= (\alpha_1, \alpha_2, \dots, \alpha_n)\\
				\beta &= (\beta_1, \beta_2, \dots, \beta_n)
			\end{split}\\
			\begin{split}
				\abs{\alpha} &= \alpha_1 + \alpha_2 + \dots + \alpha_n\\
				\abs{\beta} &= \beta_1 + \beta_2 + \dots + \beta_n
			\end{split}\\
			\partial^{\alpha}\partial^{\beta} f &= \frac{\partial^{\abs{\alpha} + \abs{\beta}} f}{\partial A_1^{\alpha_1} \partial B_1^{\beta_1} \cdots \partial A_n^{\alpha_n} \partial B_n^{\beta_n}}\\
			\begin{split}
				\mathbf{A}^\alpha &= A_1^{\alpha_1} A_2^{\alpha_2} \cdots A_n^{\alpha_n}\\
				\mathbf{B}^\beta &= B_1^{\beta_1} B_2^{\beta_2} \cdots  B_n^{\beta_n}
			\end{split}\\
			\begin{split}
				\mathbf{a}^\alpha &= \sin^{\alpha_1}(n_1\phi + \theta_1)  \sin^{\alpha_2}(n_2\phi + \theta_2) \cdots\\
				\mathbf{b}^\beta &= \cos^{\alpha_1}(n_1\phi + \psi_1)  \cos^{\alpha_2}(n_2\phi + \psi_2) \cdots
			\end{split}
		\end{align}
		This unusual double notation is required because the number of differentiations $\abs{\alpha}$ and $\abs{\beta}$ with respect to $\mathbf{A}$ and $\mathbf{B}$ has to be counted separately for Eq. \eqref{eq:partial-d-general-solution}.
		\subsection{Derivation of the residual error bound}
		Because $-1 \le \sin x \le 1$, it is clear that
		\begin{equation}
			\left | \partial^{\alpha}\partial^{\beta} f(\mathbf{0}, \mathbf{0}) \right | \le (\abs{\alpha}+\abs{\beta}-1)!
		\end{equation}
		for any angle $\phi$ (Eq. \eqref{eq:partial-d-general-solution}). Note that this is not a tight bound, because the phase relations are ignored. This leads us to an important bound of the remainder of the Taylor series.
		
		Similarly to Eq.\,\eqref{eq:mean-squared-amp}, let
		\begin{equation}
			\tilde{\mathcal{A}} = \sum_{n}\left(A_n + B_n \right) = \norm{\mathbf{A}}_1 + \norm{\mathbf{B}}_1 \,.
		\end{equation}
		
		For a single specific approximation order $k = \abs{\alpha} + \abs{\beta}$, the Taylor series contains the terms
		\begin{equation}
			T^k = \sum\limits_{\abs{\alpha} + \abs{\beta} = k}
			\frac{\partial^\alpha\partial^\beta f(\mathbf{0}, \mathbf{0})}{\alpha! \beta!} \mathbf{A}^\alpha \mathbf{B}^\beta.
		\end{equation}
		
		By virtue of the multinomial theorem and Eq.\,\eqref{eq:partial-d-general-solution} follows:
		\begin{align}
		\begin{split}
			T^k &\le \frac{( \abs{\alpha} + \abs{\beta} -1)!}{(\abs{\alpha} + \abs{\beta})!} (A_1 + B_1 + A_2 + B_2 + \dots + A_n + B_n)^{\abs{\alpha} + \abs{\beta}}\\
			&= \frac{1}{k} (A_1 + B_1 + A_2 + B_2 + \dots + A_n + B_n)^k\\
			&= \frac{\tilde{\mathcal{A}}^k}{k} \,.
			\end{split}
		\end{align}
		The full Taylor series\footnote{Note that we established earlier in section \ref{sec:taylor-approx} that $T^0 = 0$ by definition.} is therefore given by
		\begin{align}
			\sum\limits_{k=1}^\infty T_k \le \sum\limits_{k=1}^\infty \frac{\tilde{\mathcal{A}}^k}{k} = -\log(1-\tilde{\mathcal{A}}) \,,
		\end{align}
		which is the polylogarithm $\mathrm{Li}_1(\tilde{\mathcal{A}})$.
		
		We can see that this series converges for $0 \le \tilde{\mathcal{A}} < 1$. A bound for the remainder $R_k$ of the Taylor series of order $k = \abs{\alpha} + \abs{\beta}$ is then given by subtracting the relevant orders up to $k$:
		\begin{equation}
			\label{eq:taylor-bound}
			\abs{R_k(\mathbf{A}, \mathbf{B})} \le 
			-\log(1-\tilde{\mathcal{A}}) - \sum\limits_{q = 1}^{k} \frac{1}{q}{\tilde{\mathcal{A}}^q}
			\,.
		\end{equation}
		Note that for $k=0$ as well as for large amplitudes close to 	$\tilde{\mathcal{A}} \approx 1$, Eq.\,\eqref{eq:geometric-upper-bound} provides a tighter bound of the angular error, presumably due to accumulation of the error introduced by ignoring the phase shift $(\abs{\alpha}+2\abs{\beta}) \frac{\pi}{2}$ in Eq.\,\eqref{eq:partial-d-general-solution}. 
    
    \bibliography{literature}
    
\end{document}